\documentclass[conference]{IEEEtran}
\IEEEoverridecommandlockouts
\usepackage{cite}
\usepackage{amsmath,amssymb,amsfonts}
\usepackage{algorithmic}
\usepackage{graphicx}
\usepackage{textcomp}
\usepackage{xcolor}
\usepackage{hyperref}
\usepackage{tabularx}

\renewcommand{\vec}{\mathbf}

\def\BibTeX{{\rm B\kern-.05em{\sc i\kern-.025em b}\kern-.08em
    T\kern-.1667em\lower.7ex\hbox{E}\kern-.125emX}}
\begin{document}

\title{Inference and De-Noising of Non-Gaussian Particle Distribution Functions: A Generative Modeling Approach\\

\thanks{Partially Supported by Department of Energy Grant 17-SC-20-SC}
}

\author{
\IEEEauthorblockN{1\textsuperscript{st} John Donaghy}
\IEEEauthorblockA{\textit{Department of Physics} \\
\textit{University of New Hampshire}\\
Durham NH \\
0000-0002-2700-1033}
\and
\IEEEauthorblockN{2\textsuperscript{st} Kai Germaschewski}
\IEEEauthorblockA{\textit{Department of Physics} \\
\textit{University of New Hampshire}\\
Durham NH \\
0000-0002-8495-6354}
}

\maketitle

\begin{abstract}
The particle-in-cell numerical method of plasma physics balances a trade-off between computational cost and intrinsic noise.  Inference on data produced by these simulations generally consists of binning the data to recover the particle distribution function, from which physical processes may be investigated.  In addition to containing noise, the distribution function is temporally dynamic and can be non-gaussian and multi-modal, making the task of modeling it difficult.  Here we demonstrate the use of normalizing flows to learn a smooth, tractable approximation to the noisy particle distribution function.  We demonstrate that the resulting data driven likelihood conserves relevant physics and may be extended to encapsulate the temporal evolution of the distribution function.  
\end{abstract}


\section{Introduction}
Studies in computational plasma physics aim to explain experimental results and confirm theory.  Plasma theory generally takes either the kinetic or fluid approach to modeling plasma particles.

The first-principles description of a plasma is kinetic.  Kinetic theory describes the plasma as a six dimensional phase space probability distribution for each particle species.  Kinetic theory makes no assumptions regarding thermal equilibrium and thus may result in multi-modal arbitrary distributions.

In the fluid approach, it is assumed that the details of the distribution functions can be neglected and a given fluid parcel can be described by just its density, momentum and temperature. Fluid models are generally derived by marginalizing out the velocity dependence of the fully kinetic description.

Certain regimes of space and laboratory plasmas must be simulated using kinetic models, which capture all the relevant physics but are computationally more expensive.  Two numerical approaches are typically used: Continuum (Vlasov) solvers and Particle-in-Cell (PIC) methods.  
PIC codes are an example of a fully kinetic (six dimensional) solver.  PIC codes discretize the distribution function according to the vlasov equation and then sub-sample to represent regions of plasma as macroparticles. These macroparticles are advanced by fields defined by the electromagnetic Maxwell equations. This method is computationally more tractable than a direct continuum solver, however it unfortunately introduces two sources of intrinsic noise. 

The first is systemic noise inherent to the discrete plasma representation and the mapping between a discrete mesh and continuous particle positions~\cite{GERMASCHEWSKI2016305}.  While the particle’s position is represented by continuous 3D space, it must be mapped to a 3D discrete mesh where the fields live in order to interpolate the field values and advance the particle’s momentum and position.

The second source of noise is introduced in recovering the particle distribution functions from the output of the simulation.  This is known as likelihood-free inference or simulation based inference.  The samples, or particles in this case, are data generated by advancing the simulation through some number of time-steps.  The simulation with parameters $\Theta$, represents some implicit and unknown likelihood function $x \sim p(x|\Theta)$.  Traditionally this likelihood was recovered by binning the particles into histograms.  Because the simulation must make compromises on the number of particles and other numerical constants encapsulated by $\Theta$ for tractability reasons, the resulting likelihoods tend to contain noise. 

For the remainder of this work we will use the terms particle distribution function and likelihood interchangeably.  

A possible method of de-noising the likelihood lies in generative modeling.  Generative modeling has shown great success in de-noising and super resolution tasks~\cite{NIPS2012_6cdd60ea,10.5555/2999611.2999778,bigdeli2020learning,block2020generative,bd47565410ee4222abd3e1ab36efd2fb}.  Generating an accurate de-noised distribution function from PIC codes which encapsulates the underlying physics and matches the results predicted by continuum codes, would introduce a reliable method for cross-code validation as well as cut costs by allowing for inference on commodity hardware. 

\subsection{Contribution}
In this work we aim to motivate the use of robust generative modeling techniques as a novel solution to the noise inherent to the distribution functions produced by  PIC methods.  We will apply techniques from generative modeling to de-noise our non-gaussian data, performing likelihood-free inference without violating the physical constraints of the fully kinetic model.  We will then demonstrate that this technique may be expanded to encapsulate temporal dynamics.  These experiments will be used as motivation for future core-edge coupling studies mapping distributions generated from PIC codes to distributions solved by continuum codes.

\section{Background}
\subsection{Particle Distribution Function}
The baseline particle distribution function (PDF) is seven dimensional, three spatial and three velocity components plus time per ion species, $$f_s(x,y,z,u_x,u_y,u_z,t)$$  Normally, for analysis we look at a sub-domain region of the simulation to study plasma evolution.  This amounts to marginalizing the distribution function over space and taking specific time slices resulting in a multivariate gaussian where the plasma bulk flow parameterizes the means and the temperature parameterizes the covariance.  This is known as a Maxwellian distribution
\begin{equation}
f_s (u_x, u_y, u_z) = \left(\frac{m}{2 \pi kT} \right)^{3/2} \exp \left[-\frac{m(u_x^2 + u_y^2 + u_z^2)}{2kT}\right]
\end{equation}
It is important to note that this only holds true for an idealized plasma in thermal-equilibrium.  As the domain evolves through the course of a simulation various processes will cause a departure from the Maxwellian form.  The resulting PDF will be of arbitrary form and temporally dynamic, making the task of modeling density/data-driven likelihoods particularly difficult.

\subsection{Generative Modeling}
A generative model's aim is to represent a probability distribution in a tractable fashion such that it is capable of generating new samples.  Concretely, given a datapoint $\vec{x}\sim p^*(\vec{x})$, can we learn an approximation to the true distribution $p(\vec{x}) \approx p^*(\vec{x})$ such that we may generate new samples.  The likelihood of the generated samples should closely match the likelihood of the data used to train the model.  We refer to likelihoods learned from data as data-driven likelihoods (DDL).  Our data was produced by a simulation with predefined parameters which represents the implicit likelihood, so we can say we want to find the DDL which approximates $\vec{x} \sim p^*(\vec{x}|\Theta)$

Recent advances in machine learning have produced a wide variety of generative techniques.  Chief among these are variational auto-encoders (VAE), generative adversarial networks (GAN), and expectation maximization (EM) algorithms. 

The VAE is a maximum likelihood estimator that approximates the evidence by maximizing the evidence lower bound~\cite{kingma2014autoencoding}.  The core problem with this approach lies in the approximation of the posterior.  In achieving a closed form solution, one must know a-priori the posterior's functional form.  The standard approach assumes a gaussian, as such it performs poorly on multimodal or non-gaussian data.  Alternative posteriors have been proposed in the literature~\cite{dilokthanakul2017deep}, but these methods still require a-priori knowledge of the posterior's functional form.

The GAN on the other-hand, doesn't actually model the likelihood of the data.  Its goal is to trick a discriminator into believing the generated samples have been drawn from the true distribution~\cite{goodfellow2014generative}.  So while samples generated from the GAN may appear to be reflective of the simulation data, the possibility exists that we are not modeling the true likelihood.  Relying on believable but arbitrary samples leaves no guarantee that our inference would respect the physical constraints of the domain in question.  

EM algorithms performed on gaussian mixture models do well at modeling multimodal distributions, however it requires prior knowledge of the modality of the data.  As we are looking to model our particle distribution functions at an arbitrary time during the evolution of the simulation, the modality is assumed to be dynamic. 

\subsection{Normalizing Flows}
A normalizing flow describes the transformation of a probability density through a sequence of invertible mappings~\cite{dinh2015nice}.
Given data $\vec{x} \in X$, a tractable prior $\vec{z} \sim p_z(\vec{z})$, and a learnable bijective transformation $f_\theta: X \to Z$ we can apply the following change of variable formula to define a distribution on $X$.  
\begin{equation}
\log p_x(\vec{x}) = \log p_z(\vec{z}) + \log \left| \det \frac{d\vec{z}}{d\vec{x}} \right|
\end{equation}
Furthermore, defining $f$ to be a composite of a sequence of N  bijective mappings, $f\equiv f_1\circ f_2\circ \ldots\circ f_N$ allows us to say
\begin{equation}
\log p_x(\vec{x}) = \log p_z(\vec{z}) + \sum_{i=1}^N \log \left| \det \frac{\partial\vec{h_i}}{\partial\vec{h}_{i-1}} \right|
\end{equation}
where $\vec{z} = \vec{h}_N$ and $\vec{x} = \vec{h}_0$.  Optimizing on the negative log loss gives us a maximum likelihood model that allows for efficient sampling and density estimation.  What remains to be specified is the class of bijective transfomation being used.  To make this tractable, we would ideally pick a class which is easily invertible, flexible, and results in a Jacobian with a tractable determinant.  For this work we use the Masked Autoregressive Flow (MAF).  

The MAF offers a robust procedure for modeling our DDL.  As an autoregressive model it aims to construct a conditional probability distribution for each feature, where the distribution is conditioned on all previous features.  Assuming normal priors allows us to concisely say:
\begin{equation}
p(x_i|x_{1:i-1}) = \mathcal{N}(x_i|\mu_i, (\exp\alpha_i)^2)
\end{equation}
\begin{equation}
\mu_i = f_\theta(x_{1:i-1}),\ \alpha=f_\phi(x_{1:i-1})
\end{equation}
Where $f_\theta$, $f_\phi$ are arbitrary functions parameterized by neural networks.  We may generate new data as follows
\begin{equation}
x_i = z_i e^{\alpha_i} + \mu_i
\end{equation}
To ensure robust predictions we include a permutation of the features before each layer of the flow.  This class of transformation, being autoregressive, results in a lower triangular Jacobian.  It also easily extends to conditional probabilities.  For further details on MAF please see~\cite{papamakarios2018masked}.  

We can see that the normalizing flow is convenient not only because it allows us to generate samples in an interpretible manner, but gives direct access to the density, allowing us to solve the likelihood-free inference problem for the particle distribution function.  For further details on normalizing flows we refer the reader to~\cite{dinh2017density,kingma2018glow,rezende2016variational}

\section{Experiments}
The following experiments were performed with data produced by the Particle Simulation Code (PSC)~\cite{GERMASCHEWSKI2016305}.  Multi-modal and non-gaussian behavior manifests itself in our data due to excitation processes.  Particle excitation occurs through the acquisition of energy from an outside source, usually due to magnetic reconnection or collisionless shocks.  In this case, our simulation parameters are very nearly described by~\cite{Lezhnin_2021}. 

\begin{figure}
\centering
\centerline{\includegraphics[width=0.5\textwidth]{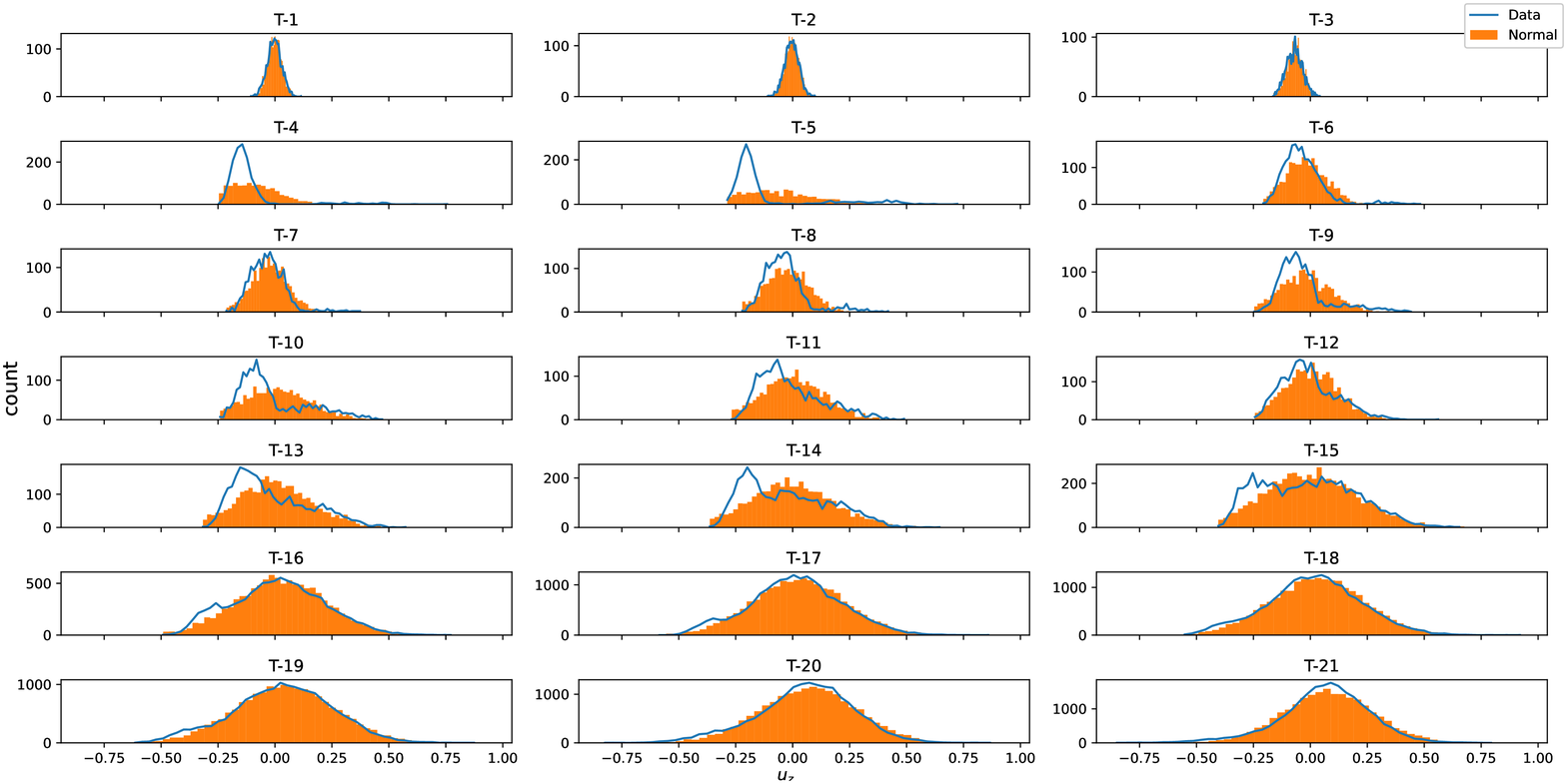}}
\caption{Temporal evolution of single feature at the same spatial region.  Inflow currents are responsible for increasing the number of particles and energization processes are responsible for driving the multi-modal evolution of the distribution.  The curve represents the particle distribution function while the solid shading is a normal distribution paramterized by our data, each containing an equal number of datapoints.} \label{fig1}
\end{figure}

Shown in Fig~\ref{fig1} is the temporal evolution of the data’s $u_z$ marginalized distribution function (not normalized).  We see that from T-4 to T-15 an energization process occurs which drives the multi-modal behavior.  Overlayed with the PDF is the normal distribution parameterized by our data’s mean and variance.

\subsection{Non-Gaussianity}
To motivate our use of the MAF we first demonstrate that our data is non-gaussian. 
There are several methods which may be used to demonstrate this, including the $t$-statistic of skewness and kurtosis, pairwise non-gaussianity of datapoints, and the Kullback-Leibler (KL) divergence.  This suite of tests outlined by Diaz Rivero~\cite{Diaz_Rivero_2020} gives us an established holistic evaluation procedure.  

\begin{figure}
\centering
\centerline{\includegraphics[width=0.5\textwidth]{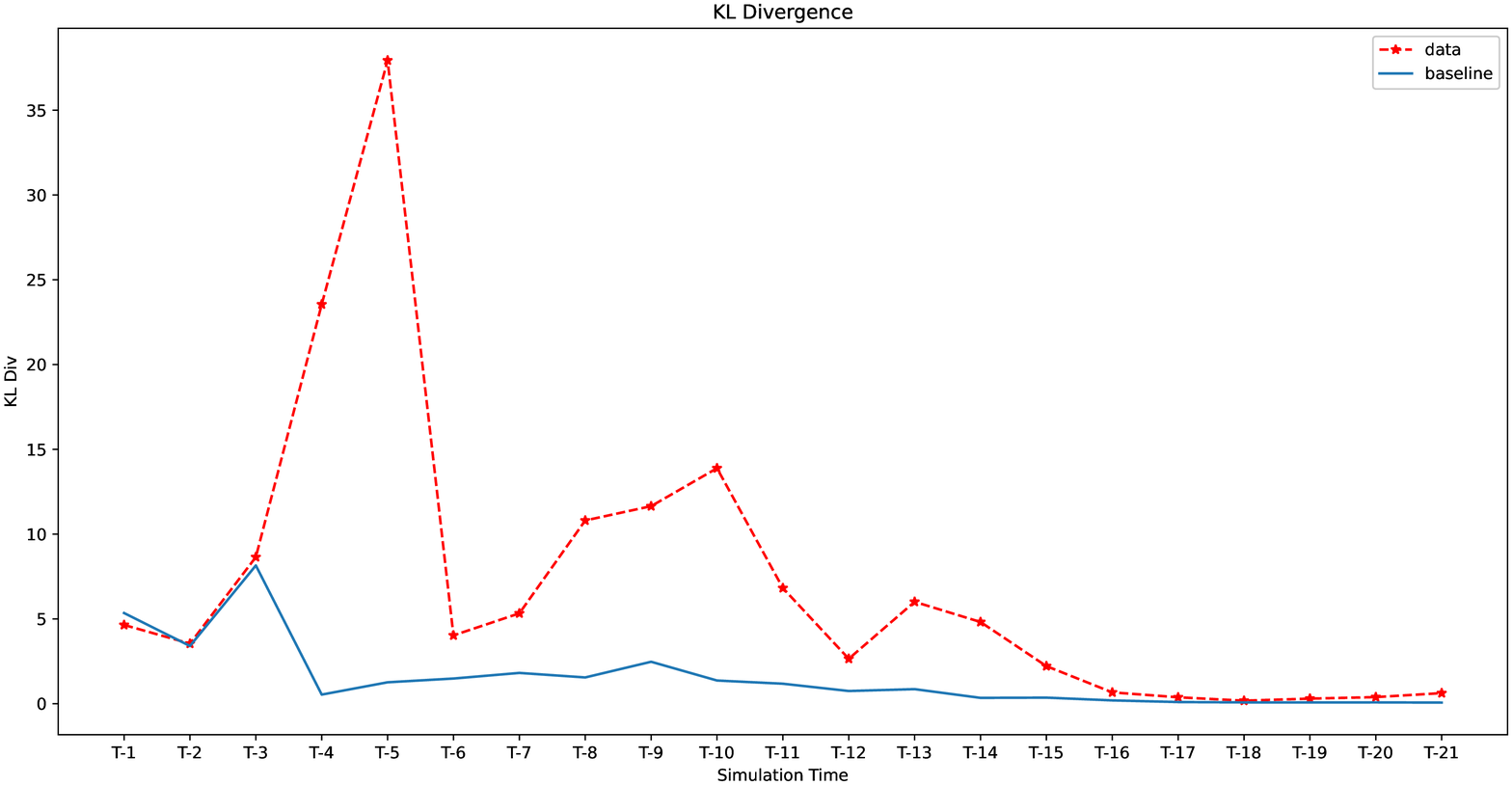}}
\caption{The Kullback-Leibler Divergence between the particle's normalized PDF and a normal distribution along with the divergence between two samples sets drawn from the normal distribution as the null hypothesis. From T-4 to T-15 there is a departure from the null hypothesis.} \label{fig2}
\end{figure}

For brevity we focus only on the Kullback-Leibler divergence test. Taking the null hypothesis to be that the our data is gaussian, we generate a normal distribution parameterized by the mean and variance of the data.  We draw two separate sample batches from the normal distribution and calculate the KL divergence between the two in order to calculate the null hypothesis.  It is well established that the KL divergence between two sample sets drawn from the same distribution will be variable on both the number of samples drawn and number of bins.  Taking both numbers to be very large we are able to minimize this variability and achieve the expected minimal distance for the baseline.  We then calculate the KL divergence between the data and a sample batch drawn from the normal distribution for comparison. Results in Fig~\ref{fig2} show that from T-4 to T-15 the KL divergence of the data is an order of magnitude greater than if the data was normally distributed, disproving the null hypothesis.  This tells us that the data is non-gaussian (non-Maxwellian) and that there are excitation processes occurring.

\subsection{Data Driven Likelihood}
Having shown the non-gaussianity of the data we can confidently state that the VAE, GAN, and EM algorithm will yield poor DDL.   
With this in mind we select the MAF as our generative model.  Nflows, built and maintained by~\cite{nflows}, is a standardized python library built on pytorch which provides a probabilistic machine learning framework.  We constructed the MAF using Nflows and trained using negative log likelihood for 1000 epochs.  The specific architecture consisted of an 8 layer flow, each layer of which contained a reverse permutation transformation and a masked affine autoregressive transformation.  The affine transformations themselves consist of a scale and shift parameter, each of which is represented by a single hidden layer neural network containing 32 nodes.  We take our base distribution to be a multivariate normal.

Training the flow using the negative log likelihood allows us to use the Adam optimizer to iteratively update the parameters of our model in an unsupervised manner.  The flow is fed simulation data which is transformed and mapped to the base distribution.  Each iterative update modifies the flow's parameters so that the likelihood of the simulation data under the base distribution after transformation is maximized.
\begin{table}
\caption{Masked Autoregressive Flow architecture}
\centering
\begin{tabularx}{\columnwidth}{|X|X|X|X|X|}
\hline
\multicolumn{5}{|c|}{MAF Hyperparameters} \\ \hline
 Layers & Permutation & Transfor-mation & hidden nodes & Base Distribution  \\ \hline
 8 & Reverse & Masked Affine Autoregressive & 32 & Multivariate Normal  \\ \hline
\end{tabularx}
\end{table}

\begin{figure}
\centering
\centerline{\includegraphics[width=0.5\textwidth]{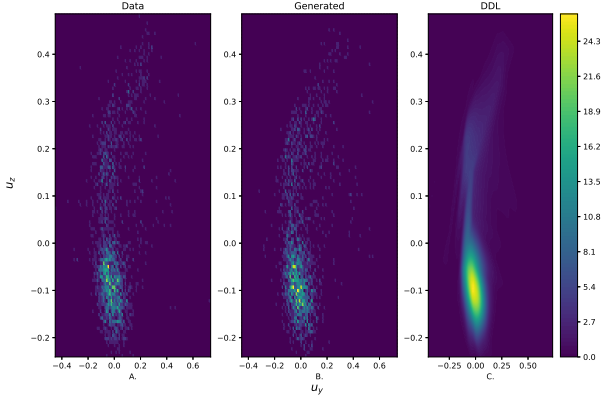}}
\caption{A. and B. normalized 2-D histograms of $u_z$-$u_y$ for particle data and generated samples respectively.  C. Data driven likelihood approximating $p^*(\vec{x}|\Theta)$} \label{fig3}
\end{figure}

Results may be seen in Fig~\ref{fig3} which show the binned distribution function of both the true data and samples generated by the model, with the smooth learned likelihood. Here we see the power of using a normalizing flow as a generative model.  By gaining direct access to the density function we are able to work with a smooth approximation to what would otherwise be a noisy distribution.  If we were to use this data to analyze kinetic processes we would traditionally use the PDF represented in frame A of Fig~\ref{fig3}.  This clearly contains noise at a level which could skew interpretation of the underlying physics.  In frame C we see the DDL learned by the model, demonstrating a dramatic noise reduction in comparison to frame A.

\subsection{Temporal Evolution}
We can leverage the versatility of the normalizing flow by taking our base distribution to be a conditional normal where we condition on simulation time.  This allows us to capture the underlying particle information at different times throughout the simulation and encapsulate that in our model.  This is powerful in that we no longer need to store terabytes of particle data, we can compress that information into the parameters of our model and perform inference from commodity hardware.   

Here we use a single layer neural network with 8 nodes and a ReLU activation to map the simulation time to the conditional parameters of our base distribution. We repeat the same training procedure as the previous section with the exception that we now use the data produced by the simulation at each interval of 1000 time-steps.  

In the framework of kinetic theory, we can use the distribution function to directly calculate the conserved physical quantities of our system.  The zeroth order moment is the number density, which may be scaled to the mass or charge density.  The first order moment gives us momentum and the second order moment kinetic energy.  

\begin{figure}
\centering
\centerline{\includegraphics[width=0.5\textwidth]{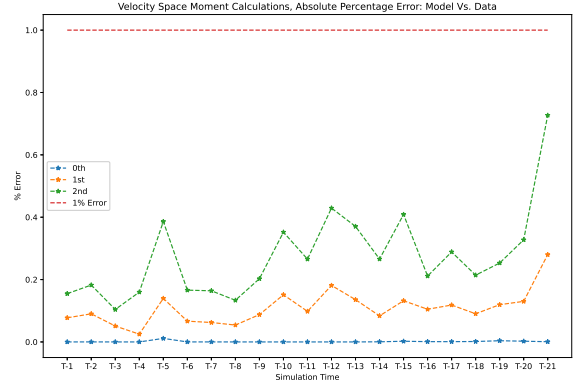}}
\caption{Zeroth, first, and second velocity moments were calculated using both the model and the true simulation data every 1000 timesteps.  The two predictions are then compared using the absolute percentage error, with results shown using 1\% error as a threshold.} \label{fig4}
\end{figure}

In Fig~\ref{fig4} we present the absolute percentage error of the zeroth, first, and second moment calculations directly between the raw data and the predictions of our model. As shown, the maximum error for the first 21,000 time-steps is always well below 1\%, demonstrating that we have compressed the temporal evolution of our simulation into our generative model without violating the physical constraints of the system.  

\section{Discussion and Conclusion}
We have shown our data to be non-gaussian and shown that we must be selective in which techniques we use to model it.  We have shown that generative modeling with normalizing flows is flexible enough to learn our PDF.  By applying the MAF to high dimensional particle data produced in PIC simulations we have successfully learned the DDL of the particles, resulting in a smooth tractable estimate of $p^*(\vec{x}|\Theta)$.  The MAF is easily extendable to conditional distributions, which allowed us to encapsulate temporal dynamics into our model and which opens up room for further studies on adaptable sub-domains.  Most importantly in modeling our data we made no assumptions as to the physical process taking place within the simulation.  Our predictions align with the simulation's results implying that we have not violated physical constraints in generating new samples.  

This presents exciting opportunities for the eXascale Computing Project Whole Device Modeling Application (WDMApp).  WDMApp aims to model plasma within the interior of a magnetic confinement fusion device known as a tokomak.  Due to the high computational cost, simulations of this nature have historically been restricted to limited volumes of the domain.  WDMApp will use the continuum code GENE to model the dense core plasma and a separate, possibly PIC code, to model the less dense edge regions~\cite{8588752}. Domain coherence requires frequent communication of the electromagnetic fields and the particle distribution function between the two codes.  The efficient transfer of this information is known as core-edge coupling and involves mapping information between the two code's disparate representations.  Coupling the codes to allow information exchange in a meaningful way is an active area of research~\cite{osti_1763841,Dominski_2018,doi:10.1063/5.0026661}.  We propose using these results as motivation for further studies incorporating generative modeling into core-edge coupling schema.

%
%
%
\bibliographystyle{splncs04}
\bibliography{mybibliography}
\end{document}